\documentclass[fleqn,twoside]{article}
\usepackage{espcrc2}

\usepackage{epsfig}
\usepackage[figuresright]{rotating}

\title{High frequency features in the 1998 outburst of 4U~1630--47}

\author{M. Klein-Wolt\address[]{Astronomical Institute Anton Pannekoek, University of Amsterdam, \\Kruislaan 403, 1098 SJ Amsterdam, The Netherlands}%
        \thanks{klein@science.uva.nl},
        J. Homan\address[]{Osservatorio Astronomico Di Brera, \\via E. Bianchi 46, 23807 Merate (LC), Italy}
		    ,~\& M. van der Klis$^{a}$%
      	}
       
\begin{document}

\begin{abstract}
We report on the detection with RXTE of variable high frequency features
in the power spectra of the black hole candidate 4U~1630--47 during its 1998 outburst. We find high frequency features during the rise and the early decay of the outburst in the range of $\sim100$--$300$ Hz. The features are usually found to be broad, and have rms amplitudes of $\sim2$--$8$\%; their frequencies are correlated with count rate and anti-correlated with spectral hardness. Also, for the first time twin high frequency features are detected simultaneously, with a ratio in frequencies consistent with 1:4. Although some of the high frequency features are less coherent and stronger than found in other black hole sources, we show that they behave very similar to the high frequency features found in for instance XTE~J1550--564.
\vspace{1pc}
\end{abstract}

\maketitle

\section{Introduction}

The soft X-ray transient 4U~1630--47 was discovered with \emph{Uhuru} \cite{jones76} but already detected with \emph{Vela} 5B in 1969 \cite{cruddace72}. There is no optical counterpart known and the best estimate we have for the distance is about 10 kpc \cite{parmar97}. The detection of X-ray absorption dips suggest the system has an inclination angle of $\sim60$--75$^{\circ}$ \cite{kuulkers98}. 4U~1630--47 is known for its regular outburst cycle: it was shown to have outbursts every $\sim600$ days (\cite{jones76} \cite{pried86}) however, recently this period changed to about 690 days \cite{kuulkers97a}. Based on the X-ray spectral and timing characteristics 4U~1630--47 has been classified as a black hole candidate \cite{white84,parmar86,kuulkers97b}, and it was pointed out that 4U~1630--47 shows similarities to the galactic sources GRO~J1655--40 and GRS~1915+105 (similar variability behaviour; \cite{kuulkers97a,kuulkers97b,kuulkers98}).

Currently, 8 black hole sources are known to show high frequency QPOs, including 4U~1630--47 \cite{remmor}. The other sources are: Cyg~X-1 \cite{pott02,nowak00}, GRO~J1655-40 \cite{rem99b,stroh01a}, XTE~J1650-500 \cite{homan03a}, XTE~J1550-564 \cite{miller01,homan01,rem02a}, GRS~1915+105 \cite{bell01,rem02b,stroh01b}, XTE~J1859+227 \cite{cui00,mark01,bpk02}, and most recently H~1743--322 (IGR/XTE~J1746--3213, \cite{homan03b}) was added to this list. In some of these cases (GRO~J1655-40, XTE~J1550-564 and GRS~1915+105) a doublet of high frequency features has been detected which seem to have 2:3 or 3:5 frequency ratios.

Here we report for the first time on the complete high frequency behaviour of 4U~1630--47 during its 1998 outburst. We show that 4U~1630--47 is capable of producing strong (2--6\% rms), both broad and narrow (Q$\sim$0.3--5.0) high frequency features, that are comparable in frequency to those found in other black holes. However, we find that 4U~1630--47 shows about a factor $3$ more power at high frequencies compared to other sources. Moreover we find, for the first time in a black hole, simultaneous high frequency features that have a frequency ratio of 1:4. The complete timing analysis of this outburst covering the general outburst behaviour, the low- and high frequency behaviour will be presented by Klein-Wolt, Homan \& van der Klis (2003, in prep.).

\section{The 1998 outburst}
\label{sec:outburst}

\subsection{Observations and data analysis}
\label{sec:data}

We analyse Proportional Counter Array (PCA) Rossi X-ray Timing Explorer (\emph{RXTE}) observations of 4U~1630--47 made between February 11 1998 and June 8 1998 and for a total of 101 observations we have about $300$ ksec of data. For the timing study we use high time resolution data modes, and power spectra are created of data segments 128s using the standard Fast Fourier Transform techniques (\cite{klis89} and references therein). Detector drop-outs are removed but no background or deadtime corrections are performed prior to the production of the power spectra. For each observation the power spectra are averaged. 

We pay special attention to the presence of possible power spectral features at high frequencies. Note that in the following we refer to the interval above $\sim20$ Hz as the high frequency, while the low frequency regime is the interval below $\sim20$ Hz. As explained in Klein-Wolt, Homan \& van der Klis (2003, in prep.), we estimate the Poisson level using the representation of Zhang \cite{zhangetal95,zhang95} with an additional shift to fit the data in a frequency range where no source power is expected (above $\sim3000$ Hz). After considerable experimentation we decided to produce the power spectra in the 7--60 keV range (channels 18--157, gain epoch 3\footnote{All energy ranges correspond to gain epoch 3}) for observations belonging to proposals 30178 and 30188, while for observations belonging to proposal 30172 the 5.5--104 keV range (channels 14--254) was used. These ranges are chosen in such a way to eliminate an instrumental effect that appeared as a broad feature at high frequencies ($\sim1000$ Hz) in our power spectra.

The power spectra are fitted with a multi-Lorentzian model (\cite{olive98}, \cite{nowak00}) using the approach of \cite{bpk02} to assign, to each component, a characteristic frequency which corresponds to the maximum in power for a Lorentzian feature in the $\nu$-${P}_\nu$ representation. We only keep those Lorentzian components in the fits whose significance based on the error in the integrated power (from 0--$\infty$) is more than $3\sigma$, or whose inclusion gives a significant ($>3\sigma$) improvement of the $\chi^{2}$ of the fit.

\subsection{General outburst behaviour}

\begin{figure}[!hb]
\epsfig{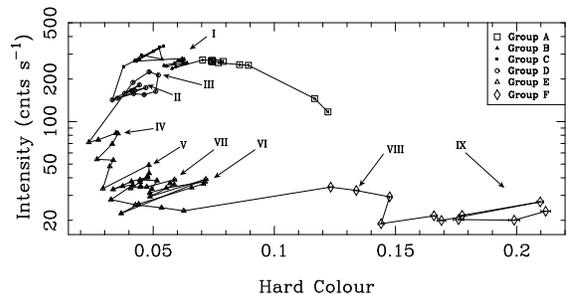}
\caption{The HID of the outburst, showing the 9 flares (I--IX). The groups (A--F) are indicated with different symbols. Hard Colour is defined as the ratio of the count rates in PCA channel bands 39--49 (15.5--19.2 keV) and 4--13 (3.0--6.2 keV), and the intensity is the 2--60 keV count rate.}
\label{fig:hid}
\end{figure}

\begin{figure}[!hb]
\epsfig{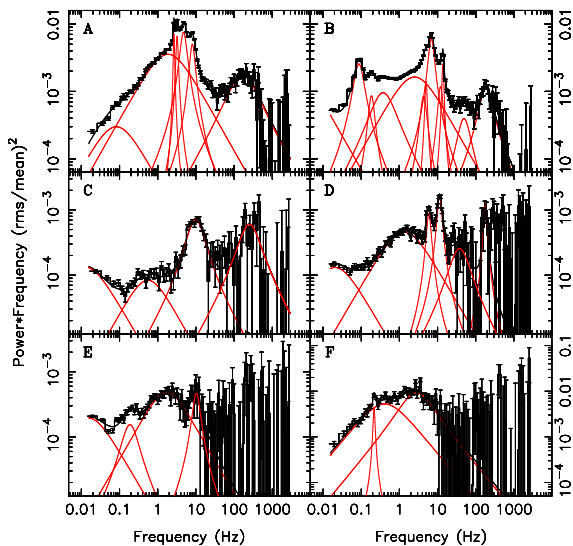}
\caption{The average power spectra of the groups A--F.}
\label{fig:groups}
\end{figure}

In Fig.~\ref{fig:hid} we show the Hardness-Intensity Diagram (HID) of the outburst. The source starts at the top-middle of the diagram, moves to the upper-left as count rate increases and hardness decreases, then moves down as the outburst declines and finally ends up at the lower-right where the count rate is low and the spectrum is hard. During the movement in the HID the source occupies different locations in the diagram that are related to specific types of timing behaviour. We find that there are six different groups (A--F) in the HID (and also in the colour-colour diagram (CD)). The average power spectra of these groups are shown in Fig.~\ref{fig:groups}; they are characteristic for the individual power spectra in each group. Note, that averaging power spectra together can result in extra features that are either due to small shifts in frequency of one component between different observations, or to the increase in significance of features that are too weak to be detected in individual observations. In group A for instance, we detect two or three features in the individual observations between 2--20 Hz, but the average spectrum in Fig.~\ref{fig:groups}a clearly shows four features. These are the result of the aforementioned frequency shifts.

The power spectra are sometimes very complex, even in the individual observations, and show features from $\sim0.1$--$300$ Hz. Notice, for instance the so called "dipping QPOs" around 0.1 Hz, already reported by \cite{diet00}, and the strong (up to $\sim7$\%) QPOs around 10 Hz (Fig.~\ref{fig:groups}). The most remarkable features, however, are the high frequency features which are detected typically above $\sim20$ Hz and are only present in groups A, B, C and D. We will discuss these in more detail in the next section.

During the outburst the source shows a number of flares, indicated with the roman numerals (I--IX) in the HID (Fig.~\ref{fig:hid}). During these flares the spectrum hardens, the rms amplitude increases and features appear or disappear in the power spectrum. In cases when the source during a flare moves in the direction of another group in the HID (for instance during flare II and III from group D to B, Fig.~\ref{fig:hid}), we find that the power spectral behaviour is very similar to that group. This, together with the fact that different groups have distinct different power spectral characteristics, suggest that the behaviour of 4U~1630--47 in the HID (and CD) is two-dimensional: the power spectral behaviour is determined by the position in the HID and not by the chronology by which the source moves through the various regions in the HID. Therefore, the power spectral behaviour depends on the two parameters that determine the position in the HID, the count rate and the hard colour (HC). This behaviour is very similar to that of XTE~J1550--564 \cite{homan01} and it suggests that both sources share a common outburst mechanism.

\section{The high frequency features}
\label{sec:hfqpo}

In Fig.~\ref{fig:groups} we show the high frequency features detected in groups A--D. In Table~\ref{table:1} we give the parameters of these components. For groups A and C we find one high frequency feature at 164 Hz and 262 Hz, respectively, that is relatively broad. In groups B and D we find two high frequency features each time, one below 100 Hz and one above. For both the twin high frequency features, we find that their ratio of frequencies is about 1:4.

In 22 of the 101 individual observations during the 1998 outburst of 4U~1630--47 we detect significant high frequency features. These features cover a range of widths with Q values of $\sim0$--$6$ and frequencies ranging from $\sim23$--$300$ Hz. For these features we find an increase in frequency during the rise of the outburst (first $\sim4$ days) which is followed by a more or less stable region during which the frequencies hardly change (remain constant at $\sim180$ Hz), and finally a decrease in frequencies during the decay of the outburst. This behaviour is reflected in a correlation with count rate and an anti-correlation with HC for all the high frequency features detected above 100 Hz. We find that the features detected below 100 Hz do not seem to follow that same relation. Note, that the same correlations between count rate and HC are also found for the features detected in the average power spectra. Finally, also in five of the individual observations in groups A, B, C and D we find a feature below 100 Hz. However, only on one of these occasions another feature above 100 Hz is detected: again we find two features (at 42 and 170 Hz) with a ratio in frequencies of 1:4.

\begin{table}[!htb]
\caption{High Frequency features in the average power spectra of groups A--F. In the last column we also give the significance of the detected features.}
\label{table:1}
\begin{tabular}{cccc}
 & $\nu_{max}$(Hz)	&Q	&rms (\%, $\sigma$) \\
\hline
A & 164.3$\pm$9.8 &0.33$\pm$0.07 &6.2$\pm$1.5 (17.7)\\
B & 49.3$\pm$3.0  &1.2$\pm$0.26  &2.1$\pm$0.9 (5.9)\\
  & 187.4$\pm$6.5 &1.09$\pm$0.19 &3.6$\pm$1.1 (10.3)	 \\
C & 262.2$\pm$45.5 &0.56$\pm$0.28	&3.3$\pm$1.5 (4.9) \\
D & 38.06$\pm$7.3 &0.67$\pm$0.37	&2.1$\pm$1.1 (4.0)	\\
 & 179.3$\pm$5.7  &4.6$\pm$2.3 &2.0$\pm$1.1 (3.8)	 \\
E & - & - &2.3~$^{1)}$   \\
F & - & - &2.9~~~   \\
\hline
\end{tabular}
~\\
\footnotesize{1)~95\% upper limit for Q=12, 50--1000 Hz range.}\\
\end{table}

\begin{figure}[!hbt]
\epsfig{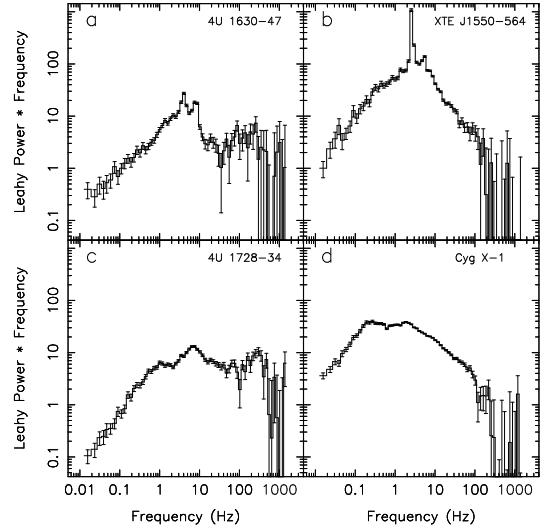}
\caption[]{The power spectra of 4U~1630--47, XTE~J1550--564, Cyg X-1 and the neutron star source 4U~1728--34. The observations are selected to have similar count rates, settings, and observation periods. The poisson level subtraction method is similar for all the sources.}
\label{fig:bhc}
\end{figure}

From Table~\ref{table:1} and Fig.~\ref{fig:groups} it is clear that the high frequency features detected in 4U~1630--47 are usually very broad and strong: we find rms amplitudes from $\sim2$--$7$ \% in both the average and individual observations. We compare the average power in the high frequency range (100--3000 Hz) with that of two black hole sources, Cyg X-1 and XTE~J1550--564 and two neutron star sources, 4U~1608--522 and 4U~1728--34. We find that on average 4U~1630--47 shows $\sim3$ times more power compared to the black hole sources and about 3 times less power compared to the neutron star sources. Also, when comparing the shape of the power spectra the difference between 4U~1630--47 and other black hole sources is evident. From Fig.~\ref{fig:bhc} it is clear that for the two black hole sources (Cyg X-1 and XTE~J1550--564) the power spectrum \emph{declines} above $\sim100$ Hz, while for both 4U~1630--47 and the neutron star 4U~1728--34 the power \emph{increases}. This clearly shows that, like neutron stars this black hole candidate, too, can show relatively high power levels at high frequencies.

\section{Discussion and conclusions}
\label{sec:disc}

\begin{figure}[!hbtp]
\begin{center}
\epsfig{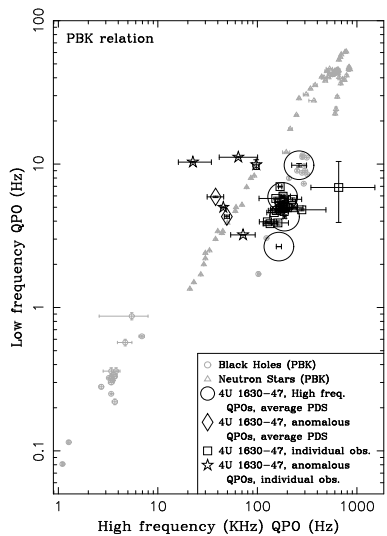}
\caption[]{The PBK relation with the points for 4U~1630--47 overplotted.}
\label{fig:pbk}
\end{center}
\end{figure}

In Fig.~\ref{fig:pbk} we show the Psaltis-Belloni-van der Klis relation (PBK relation) for a number of black hole sources and neutron star sources (taken from \cite{bpk02}). For the lower frequency QPO of 4U~1630--47 we always take the first Lorentzian of the complex of features around $10$ Hz, and we plot both the points for the individual and average power spectra. First of all, it is clear from this figure that 4U~1630--47 shows a correlation between the low frequency features and the high frequency features: 4U~1630--47 shows the same trend as the original PBK relation and falls on the lower PBK track. Notice that this is exactly at the same position as XTE~J1550--564, indicating that both sources show features at similar frequencies. Secondly, we find that the high frequency features below 100 Hz (the big stars and diamonds in Fig.~\ref{fig:pbk}) do not fall on the PBK relation, nor do they show any consistent relation between the low and high frequency features. This, together with the fact that they also do not follow the same relations with HC and count rate as the features detected above 100 Hz and the fact that they are about a factor 2 lower in frequency, strongly suggest that the features below 100 Hz are different from the other high frequency features and might even have a different origin.

The agreement of 4U~1630--47 with the PBK relation raises the question if the high frequency features in this source are related to the kHz QPOs found in the neutron star sources? Although the relation between NS and BH features over a large range in frequencies shown by the PBK relation in itself suggests a similarity between these two types of sources, we would like to stress that the most important results that can be deduced from this relation are: \emph{1) the correlation between the low and the high frequency features (PBK relation) for 4U~1630--47}; and \emph{2) the overlap in frequencies between 4U~1630--47 and the black hole source XTE~J1150-564}. Both these results clearly suggest that 4U~1630--47 can behave very similar to other black hole sources, showing similar frequencies and relations between them. 

It is for the first time for a black hole X-ray binary system that twin high frequency features are detected in the same energy band, with a frequency ratio of 1:4, clearly different from the more common 2:3 or 3:5 ratios (see for instance \cite{rem02c}). The fact that the ratio is different in the case of 4U~1630--47 might be related to the fact that the lower of the two high frequency features seems to be different, as explained above. However, during the outburst of other black hole sources similar features are found. For instance during the 2001 outburst of XTE~J1650-500 a feature around 50 Hz is detected \cite{homan03a,kalemci02}. This feature is simultaneously present with features at higher frequencies (at twice or three times the frequency) and is broad but much stronger compared to the feature at higher frequencies \cite{homan03a}. Also, similar features are reported for Cyg X-1, GX~339-4 and XTE~J1118-480. It is not clear whether these features are all the same, however, they might be related to the Hecto Hz QPOs found in both neutron star and black hole sources (see \cite{straaten02} and references therein); the anomalous high frequency features found in 4U~1630--47 have similar frequencies, however, they are not as stable in frequency as the normal Hecto Hz QPOs.

Note that, although the 1:4 ratio is unconventional, it follows for instance from the discoseismic model (\cite{wag01} and references therein) as a combination between a g- and c-mode oscillation or as a non-axisymmetric oscillation between two g-modes \cite{wag01}. Also, in the model by Abromowicz and Klu\'{z}niak it is perhaps possible to get a 1:4 ratio by putting the ratio between the relativistic epicyclic frequency and the relativistic Keplerian frequency equal to 1:4. These models do not exclusively predict the ratio of the frequencies, hence the ratio (2:3, 3:5 or 1:4) might be determined by the characteristics of the individual sources.

In conclusion, we find that during the 1998 outburst of 4U~1630--47 broad, strong high frequency features are present, which are comparable in frequency to the high frequency features detected in XTE~J1550--564. We find that on average 4U~1630--47 shows about 3 times more power in the high frequency regime compared to other black hole sources, however, also about 3 times less power compared to neutron star sources. Also, the overall behaviour of 4U~1630--47 is similar to that of XTE~J1550--564 and suggests a two-dimensional behaviour in the HID, which depends on both the count rate and the HC. Based on the behaviour described here we suggest that identifying black hole and neutron star sources based on the amount of power at high frequencies, as suggested by \cite{sun00}, is not straightforward: 4U~1630--47 is capable of showing an increase in power above $\sim100$ Hz, like the neutron star sources. In fact, our findings seem to suggest that black holes and neutron stars can show very similar behaviour, indicating that the timing properties are mainly determined by processes that take place in the accretion disc and are less dependent on the presence of a surface, boundary layer, or strong magnetic field.

{}

\end{document}